# QUANTUM EFFECTS IN THERMAL CONDUCTIVITY OF SOLID KRYPTON - METHANE SOLUTIONS

## A. I. Krivchikov, B. Ya. Gorodilov, V. G. Manzhelii and V. V. Dudkin

Institute for Low Temperature Physics & Engineering NASU, Kharkov 61103, Ukraine
Email: krivchikov@ilt.kharkov.ua

The dynamic interaction of a quantum rotor with its crystalline environment has been studied by measurement of the thermal conductivity of solid $Kr_{1-c}(CH_4)_c$ solutions at $c = 0.05$-$0.75$ in the temperature region from 2 up to 40K. The thermal resistance of the solutions was mainly determined by the resonance scattering of phonons by $CH_4$ molecules with the nuclear spin I=1 (the nuclear spin of T-species). The influence of the nuclear spin conversion on the temperature dependence of the thermal conductivity $\kappa(T)$ was found: a clearly defined minimum on $\kappa(T)$, its temperature position depending on the $CH_4$ concentration. It was shown that the anisotropy molecular field not increase monotonously with the $CH_4$ concentration. A compensation effect in the mutual orientation arrangement of the neighboring rotors is observed at $c > 0.5$.

The temperature dependence of $Kr_{1-c}(CH_4)_c$ is described within the Debye model of thermal conductivity taking into account the lower limit of the phonon mean free path. The anomalous temperature dependence of the thermal resistance shows the evolution of the phonon-rotation coupling at varying temperature. It increases strongly when the character of $CH_4$ rotation changes from the quantum at low temperatures to classical at high temperatures.

Also, a jump of thermal conductivity (a sharp increase in $\kappa(T)$ within a narrow temperature range) was observed, whose position varies from 9.7 K to 8.4 K when the $CH_4$ concentration changes from 0.25 to 0.45.

Owing to the high symmetry of the $CH_4$ molecule and to the arrangement of light H atoms at its periphery, the rotation of $CH_4$ molecules in some condensed media is nearly free even at low temperatures. In this case the energy spectrum of rotation is essentially dependent on the total nuclear spin of the rotating molecules [1]. The equilibrium concentration of the three possible nuclear spin species of $CH_4$, namely, the A, T, and E species with the total nuclear spins of the protons I = 2, 1, 0, is determined by the temperature and by the symmetry of the potential field in which the $CH_4$ molecule appears in the condensed medium. The relaxation time taken to bring the spin species to equilibrium concentration increases with decreasing temperature [2]. The experimental investigation at low temperatures should therefore take into account the real concentration of the A-, T- and E- species and the rate of their mutual transformation (the rate of spin conversion). The dynamics of the rotational motion and spin conversion were studied in detail in different phases of solid $CH_4$ [3, 4] and in nonconcentrated solid solutions of $CH_4$ in Ar, Kr and Xe crystals [5-11].

A freely rotating molecule does not interact with the surroundings. In the real situation however even weak anisotropic interactions are important when we consider certain properties of condensed phases containing rotating molecules [12]. The rotor-lattice interaction in solids is referred to as a phonon-rotation coupling (PRC). It is precisely the PRC that is responsible for the equilibrium of the translational lattice vibrations and the rotation of the molecules. The PRC can in particular affect the thermal conductivity [13-15]. In turn, the thermal conductivity can be used as a tool to study the PRC.

The solid $Kr_{1-c}(CH_4)_c$ solutions are the most suitable objects to investigate the behavior of weakly hindered rotors in the crystal. Because of the spherical symmetry of Kr atoms and close Lennard-Jones potential parameters of methane and krypton, the rotation of $CH_4$ molecules is weakly hindered; no significant dilatation occurs in the solid solution lattice, and the mutual solubility of the components in the solid phase is high ($0 < c < 80\%$). The high symmetry of the solution FCC lattice makes the interpretation of the results much easier. This system permits us to investigate both the PRC effect on the thermal conductivity and the anisotopic (noncentral) interaction within the rotor system which manifests itself in the thermal conductivity.

As previous studies [15] of the thermal conductivity of nonconcentrated $CH_4$-Kr solutions ($c \leq 5\%$) showed, the noninteracting T-species molecules are centers of strong resonance scattering of phonons. The conversion effect on the temperature dependence of the thermal conductivity of the solid solution was also examined.

The goal of this study is to investigate the PRC effect on the thermal conductivity of concentrated solid $CH_4$-Kr solutions in which the $CH_4$ subsystem can be considered as an ensemble of rotors interacting with one another and with the translational lattice vibrations.

The thermal conductivity of the solid $Kr_{1-c}(CH_4)_c$ ($c = 0.05$, 0.10, 0.25, 0.45, 0.75) was measured in the interval 2÷40 K using the steady-state technique. The measuring cell configuration and the technique of thermal conductivity measurement are described in [16]. The samples were grown in a cylindrical stainless steel cell 38 mm long and 4.5 mm in inner diameter by crystallization of the liquid solution at the equilibrium vapor pressure. The growth rate was 0.07 mm/min. The rate of cooling the sample to $T = 40$ K at which the measurement of the temperature dependence of thermal conductivity was started was 0.15 K/min. On cooling the sample, the temperature gradient 0.18 K/min along the cell was kept. The technique of sample preparation with the smallest possible contents of structural defects (grain boundaries and dislocations) was optimized in the process of growing polycrystalline samples and measurement of the thermal conductivity of pure krypton. The quality of the Kr sample could be judged from the value of the phonon maximum in the thermal conductivity. In our experiment we used krypton of natural isotopic composition, its purity being 99.94%. The Kr gas contained the following impurities: $N_2 - 0.046\%$, $Ar - 0.012\%$ and $O_2 \leq 0.0005\%$. The chemical purity of methane was 99.99%. $CH_4$ contained 0.005% $N_2$. The $O_2$ impurity was below 0.0001%. The absolute error of the thermal conductivity coefficient was within 10% below 15 K and 20% above 15 K. Temperature dependence of the thermal conductivity $\kappa(T)$ at $c = 0.05$ was measured additionally to test our previous published results [15].

The measured thermal conductivities of the $Kr_{1-c}(CH_4)_c$ solutions are shown in Fig. 1. The difference between our results and [15] is within the experimental error. Fig. 2 shows good agreement of our $\kappa(T)$ curves at $c = 0.75$ and literature data on the thermal conductivity of this solution [17] at $c = 0.66$ in the temperature interval, where the $\kappa(T)$ dependence tends to the curve taking into account the lower limit of the phonon mean free path [18].

The $\kappa(T)$ dependences taken at different $CH_4$ concentrations have two distinct features: i) a minimum in the thermal conductivity curve in the interval $4 \div 10$ K and ii) a sharp rise (jump) of the curve to the right of the temperature minimum. The feature positions on the temperature scale are dependent on the $CH_4$ concentration (see Table 1). The temperatures of the minimum ($T_{min}$) and the jump ($T_a$) decrease as the concentration is increasing to 0.45 and then increase on a further concentration growth. The minimum in the $\kappa(T)$ curve is a manifestation of resonance phonon scattering by rotational excitations of the $T-CH_4$ molecules.

The results obtained show that on a further increase in the concentration, the dip of $\kappa(T)$ curve observed in dilute solutions [15], in particular at $c = 0.05$, transforms into a minimum in concentrated solutions.

In the general case, the thermal conductivity of the $Kr_{1-c}(CH_4)_c$ solution is determined by the spin conversion and some processes of phonon scattering: i) the PRC between the transnational lattice vibrations and the $CH_4$ rotor ensemble; ii) the Rayleigh scattering due to the different Kr and $CH_4$ masses; iii) phonon-phonon scattering, iv) scattering by structural defects. Among the above mechanisms, the PRC scattering is the most difficult to describe theoretically. Below we present the procedure of separating the PRC-related contribution from the total thermal resistance of the solution.

The basic mechanisms of phonon scattering in inert gas crystals of Ar, Kr, Xe are well known and commonly accepted expressions describing them are available. The temperature dependence of thermal conductivity can usually be described by the Debye model for an isotropic solid, which ignores the difference between the phonon modes of different polarizations:

$$\kappa(T) = \frac{k_B^4 T^3}{2\pi^2 \hbar^3 s} \int_0^{\Theta/T} (\tau(x) + \frac{\pi}{k_B T x} \hbar) \frac{x^4 e^x}{(1-e^x)^2} dx; \quad (1)$$

where $k_B$ is the Boltzmann constant, $\hbar$ is the Planck constant, $\Theta$ is the Debye temperature, $s$ is the mean sound velocity, $x = \hbar\omega / k_B T$, $\tau(x)$ is the effective relaxation time of the phonons involved in scattering. The term $\pi/\omega$ is introduced into Eq.(1) to limit the thermal conductivity at high temperatures when the wavelength of the phonons becomes equal to half the lattice parameter (the limit was proposed by Cahill and Pohl [18]). The normal phonon-phonon processes in Kr, $CH_4$ and their solutions are not intensive and can therefore be ignored [12,17,19]. The inverse relaxation time (relaxation rate) $\tau^{-1}(\omega,T)$ is a sum of relaxation times for all resistive processes of phonon scattering: $\tau_U^{-1}$ (three-phonon $U$-processes), $\tau_B^{-1}$ (by the boundaries), $\tau_{dis}^{-1}(\omega)$ (by dislocations), $\tau_{Rayleigh}^{-1}(\omega)$ (by point defects) and $\tau_{rot}^{-1}(\omega,T)$ (by the rotation states of the $CH_4$ molecule):

$$\tau^{-1}(\omega,T) = \tau_U^{-1}(\omega,T) + \tau_B^{-1} + \tau_{dis}^{-1}(\omega) + \tau_{Rayleigh}^{-1}(\omega) + \tau_{rot}^{-1}(\omega,T) .$$

For the $U$-processes the relaxation rate is:

$$\tau_U^{-1}(\omega,T) = A\,\omega^2\,T\,exp(-b/T),$$

where $A \approx \hbar\gamma^2/(Ms^2\Theta)$, $\gamma$ is Grüneisen parameter, $M$ is the mean mass of the particles of the substance. The fitting parameters $A = 4.41 \cdot 10^{-16}$ sec/K, $b=16$ K were used for pure Kr [15]. $M, s, b$ and the relative variation of $A$ as a function of the $CH_4$ concentration are given in Table 2.

For boundary scattering the relaxation rate is:

$$\tau_B^{-1} = s\,/\,l,$$

where $l$ is the mean free phonon path.

For scattering at dislocations the relaxation rate is:

$$\tau_{dis}^{-1}(\omega) = D\,\omega,$$

where $D$ is the parameter dependent on the dislocation density. The fitting parameters for the above three mechanisms were found using the measurement data on the thermal conductivity of pure Kr [15]. The variation of these magnitudes was assumed negligible even in concentrated $Kr_{1-c}(CH_4)_c$ solutions.

Since the masses of Kr atoms and $CH_4$ molecules are different, the relaxation rate for scattering by point defects is expressed as:

$$\tau_{Rayleigh}^{-1}(\omega) = \frac{\Gamma\Omega_0}{4\pi s^3}\omega^4 ,$$

where $\Gamma = c\,(1-c)\,(\Delta M/M)^2$, $\Delta M$ is the mass difference between the pure components of the solution, $\Omega_0$ is the unit cell volume. The $\Gamma$-values are presented in Table 2. The changes in the force constants and the dilatation near the impurity center were neglected because the interaction parameters are close for Kr atoms and $CH_4$ molecules [1].

There is no commonly accepted expression for the relaxation rate $\tau_{rot}^{-1}$ describing the PRC mechanism. The general PRC regularities can be established through separation from the total thermal resistance of the contributions corresponding to different mechanisms of scattering. The excess thermal resistance in $Kr_{1-c}(CH_4)_c$ can be found from experimental thermal conductivities of the solution and pure Kr as $\Delta W(T) = W(T) - W_{Kr}(T)$, где $W(T) \equiv 1/\kappa(T)$. To illustrate the separation of the thermal resistance contributions, Fig. 3 shows the temperature dependence of $\Delta W/W_{Kr}$ for $c = 0.45$. There is a considerable contribution of the $CH_4$ molecules to the total thermal resistance. At helium temperatures $\Delta W$ is over an order of magnitude higher than $W_{Kr}$. The ratio $\Delta W/W_{Kr}$ decreases with increasing temperature. It is however impossible to separate the PRC-induced contribution $\Delta W_{rot}$ directly from the excess thermal resistance. This can be done using the expression

$$\Delta W_{rot}(T) = W(T) - W_{calc}(T),$$

where $W_{calc}(T)$ is the dependence obtained by Eq.(1) taking to account all the scattering mechanisms included into the sum of Eq.(2) except for the PRC mechanism.

The dependence $W_{calc}(T)$ was calculated taking into account that the Debye temperatures $\Theta$ of the pure components Kr and $CH_4$ differ considerably (see Table 2). As the concentration changes from 0 to 1, the Debye temperature increases from 71.7 K to 140 K [20], while the mean mass $\bar{M}$ decreases from 83.8 to 16. For a concentrated solution, $\Theta$ can be estimated roughly as a function of $CH_4$ concentration [21]:

$$\Theta(c) = \Theta_{Kr}\left(\frac{V_{Kr}}{V(c)}\right)^d \sqrt{\frac{M_{Kr}}{M(c)}} , \text{ (3)}$$

where $V$ is the molar volume, $M(c) = (1-c)M_{Kr} + c\ M_{CH4}$ and $V(c) = (1-c)V_{Kr} + c\ V_{CH4}$, $d$ is the parameter equal to the Grüneisen constant for isotopic solutions. The $d$-value for $Kr_{1-c}(CH_4)_c$ was found by substitution of $\Theta(c)$, $V(c)$, $M(c)$ of Eq. (3) with the corresponding values obtained for pure solid $CH_4$. The sound velocity $s$ was calculated using $\Theta(c)$, $V(c)$:

$$s = \frac{k_B}{\hbar(6\pi^2 N_A)^{1/3}} \Theta(c) V(c)^{1/3} ,$$

where $N_A$ is the Avogadro number. $V$, $\Theta$, and $s$ for different $c$ are listed in Table 2.

The curves $\Delta W_{rot}(T)$ describing for the PRC-induced thermal resistance are shown in Fig. 4 for different $CH_4$ concentrations.

The concentration dependence of $\Delta W_{rot}$ is nonmonotonic. A change in the concentration from 0.05 to 0.45 leads to an increase in the thermal resistance. A further increase in $c$ from 0.45 to 0.75 has practically no effect on $\Delta W_{rot}$.

The nonmonotonic dependence $\Delta W_{rot}(T)$ describes the PRC evolution in the system of interacting rotors with temperature. The dependence $\Delta W_{rot}(T)$ has two maxima, a low temperature maximum in the temperature region of the orientational disordered phase with a frozen orientational disorder and a maximum in the dynamic disorder phase [6]. In the whole interval of temperatures $\Delta W_{rot}(T)$ is determined by resonance phonon scattering at the rotational excitations of the T-$CH_4$ molecules [15]. The decrease in the thermal resistance $\Delta W_{rot}$ at decreasing temperature observed to the left of the maximum occurs because at decreasing temperature the number of T-molecules decreases due to the T-A spin conversion. The conversion is more intensive in the system of interacting rotors [11], where it leads to an equilibrium distribution of the $CH_4$ species in the studied temperature interval for $c > 10\%$. In the system of noninteracting rotors at $c < 10\%$ and T< 5K, the curve $\Delta W_{rot}(T)$ is observed to rise the temperature decreases. The increase can be explained assuming that the T-$CH_4$ concentration exceeds the equilibrium value and does not change with temperature.

In the low temperature interval (2÷10K) the molecules in the solutions studied have a well-structured rotational spectrum with a slightly broadened of the energy levels for both noninteracting molecules and the interacting rotor systems [6]. The finite width of the energy levels is determined by the internal field surrounding the molecule and by the interaction with the phonon spectrum. In the interval above 10 K the broadening of the energy levels is due to the enhanced interaction between the molecule and the lattice vibrations. The anomalous behavior of $\Delta W_{rot}(T)$ in the temperature region, where the energy levels are broadened is the evidence of the PRC enhancement with increasing temperature.

Note another feature of the $\Delta W_{rot}(T)$ behavior, namely, the jump in a narrow temperature interval (a 25% increase) at decreasing temperature. The relative change in the thermal resistance at the temperature of the jump is independent of the $CH_4$ concentration.

A possible reason for the jump may be connected with the cooperative phenomena in the system of interacting rotors. For example, in a certain region of $CH_4$ concentrations and temperatures, the behavior of the system of interacting rotors can be influenced by its so called return character [22]: when the temperature increases, the disordered system changes into the ordered state; however, on a further rise of temperature the system comes back to the disordered state. Thus, the system undergoes two phase transitions at increasing temperature: a low-temperature orientational disorder-order (quantum nature) transition and the high-temperature change from the ordered state into the disordered phase of classical nature. The behavior of the system of interacting $CH_4$ molecules in $Kr_{1-c}(CH_4)_c$ is determined by the octupole - octupole interaction which becomes more intensive with rising temperature because the number of T-$CH_4$ molecules increases (the T-$CH_4$ molecule has the larger octupole moment) and then becomes weaker as the $CH_4$ molecules are being partially substituted with the Kr atoms.

The authors would like to thank R. Pohl, W. Press, G. Weiss, A. Jezhowski and B. Donilchenko for helpful discussions.

Table 1. Parameters, characterizing the behavior of the thermal conductivity of $Kr_{1-c} (CH_4)_c$. Symbols: $T_{min},\ T_a$ are the temperatures of minimum and jump, accordingly, $K_{min}$ is the value of thermal conductivity at $T_{min}$, $\Delta K_a$ is the change of the thermal conductivity and $\Delta W_{rot}/\Delta W^+_{rot}$ is the change of relative thermal resistivity at $T_a$ ($\Delta W_{rot}$ and $\Delta W^+_{rot}$ are the values of $\Delta W_{rot}$, determined by extrapolation of $\Delta W_{rot}(T)$ from the left and to the right relatively (from) of $T_a$).

| $c$, % | $T_{min}$, K | $K_{min}$, W/(m K) | $T_a$, K | $\Delta K_a$, W/(m K) | $\Delta W_{rot}/\Delta W^+_{rot}$ |
|---|---|---|---|---|---|
| 5 | $8.3 \pm 0.3$ | $0.31 \pm 0.02$ | --- | --- | --- |
| 10 | $7.6 \pm 0.3$ | $0.133 \pm 0.004$ | --- | --- | --- |
| 25 | $6.9 \pm 0.3$ | $0.081 \pm 0.003$ | $9.7 \pm 0.2$ | $0.025 \pm 0.006$ | 1.25 |
| 45 | $4.7 \pm 0.2$ | $0.043 \pm 0.002$ | $8.2 \pm 0.2$ | $0.015 \pm 0.004$ | 1.25 |
| 75 | $6.5 \pm 0.2$ | $0.043 \pm 0.002$ | $8.5 \pm 0.2$ | $0.016 \pm 0.002$ | 1.25 |

Table 2. The values using to calculate the thermal conductivity by the Debye model. Parameters $A(c)/A(0)$, and $b$ phonon-phonon mechanism and parameter $\Gamma$ for Rayleigh scattering mechanism .

| $c$ | $\overline{M}$, g/mol | $V$, sm³/mol | $\Theta$, K | $s$, m/s | $\Gamma$, $10^5$ | $A(c)/A(0)$ | $b$ |
|------|------|------|------|------|------|------|------|
| 0 | 83.8 | 27.13 | 71.7 | 856.9 | 0 | 1 | 16 |
| 0.05 | 80.4 | 27.41 | 72.6 | 870.6 | 0.546 | 0.995 | 16 |
| 0.10 | 77.0 | 27.69 | 73.5 | 884.4 | 1.084 | 0.990 | 16 |
| 0.25 | 66.8 | 28.54 | 76.9 | 934.7 | 2.616 | 0.985 | 16 |
| 0.45 | 53.3 | 29.66 | 83.4 | 1027 | 4.265 | 1.001 | 16 |
| 0.75 | 33.0 | 31.34 | 101 | 1269 | 4.726 | 1.107 | 16 |

Fig. 1. The experimental dates of thermal conductivity of solution $Kr_{(1-c)}CH_{4c}$ for various concentrations of $CH_4$ as a function of temperature:  a) $c = 0.05$; b) – 0.10; d) – 0.25; e) – 0.75. Symbols: ↑ and ↓- temperature values of minimum and jump of thermal conductivity curves, accordingly.

Fig. 2. Thermal conductivity $Kr_{(1-c)}CH_{4c}$ with comparison with literature data:  $c = 0$ (pure Kr) and $c = 0.05$ [15]); $c = 0.66$ [17].  The present data- $c = 0.75$. The curve is calculated dependence of the limiting minimum value of $\kappa_{min}$ $(T)$ [18] for case of pure Kr.

Fig.3.  The excess thermal resistance $\Delta W(T)$ =$W(T)$-$W_{Kr}(T)$  for $c = 0.45$,  divided on thermal resistance of pure Kr as a function of temperature.

Fig. 4. The thermal resistance $\Delta W_{rot}$ due to PRC mechanism in $Kr_{(1-c)}CH_{4c}$  solution for various concentrations of $CH_4$ as a function of temperature: a) $c = 0.05$; b) – 0.10; d) – 0.25; e) – 0.75. The arrows ↑ and ↓ indicate the temperature of the maximum and the jump of thermal resistance curve. The curve is the dependence $\Delta W \sim 1/T^{1/2}$.

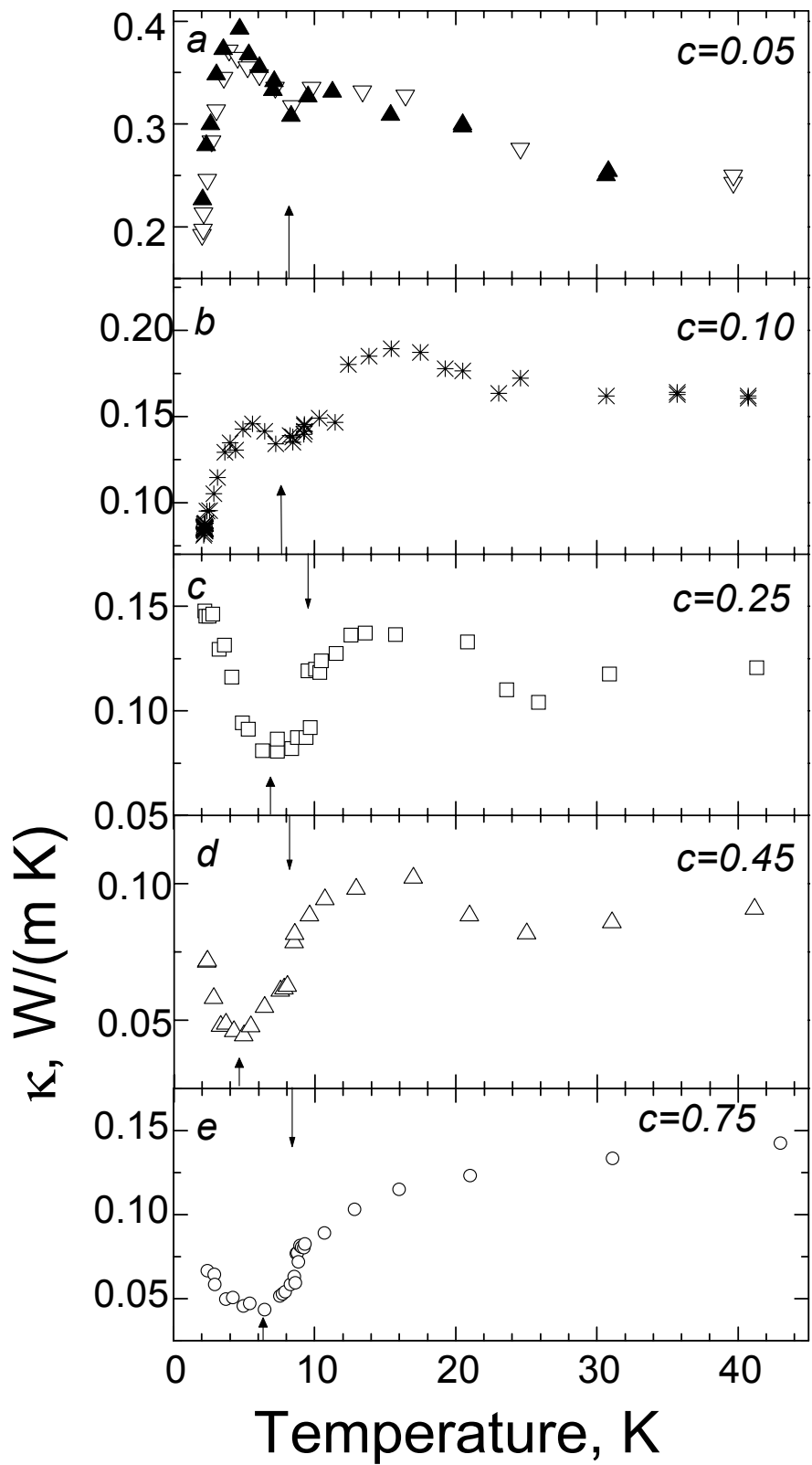

Fig.1

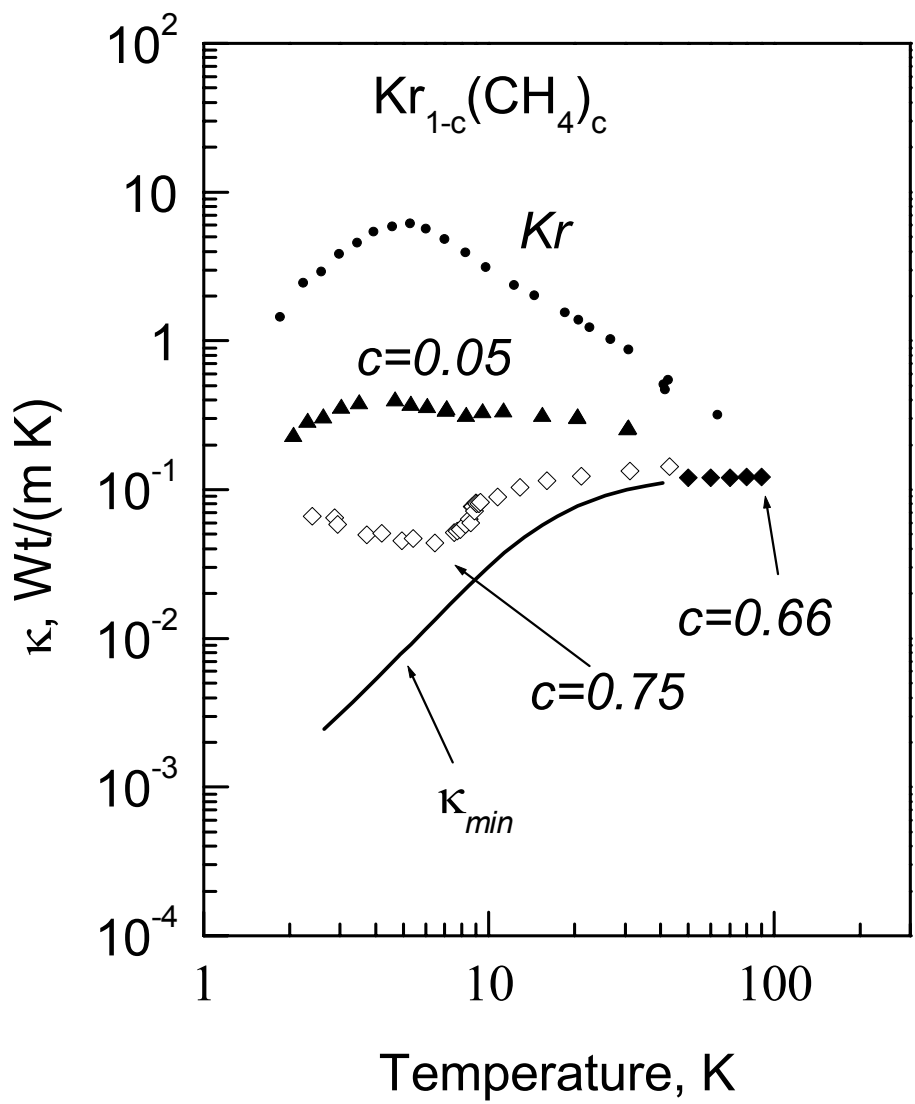

Fig.2

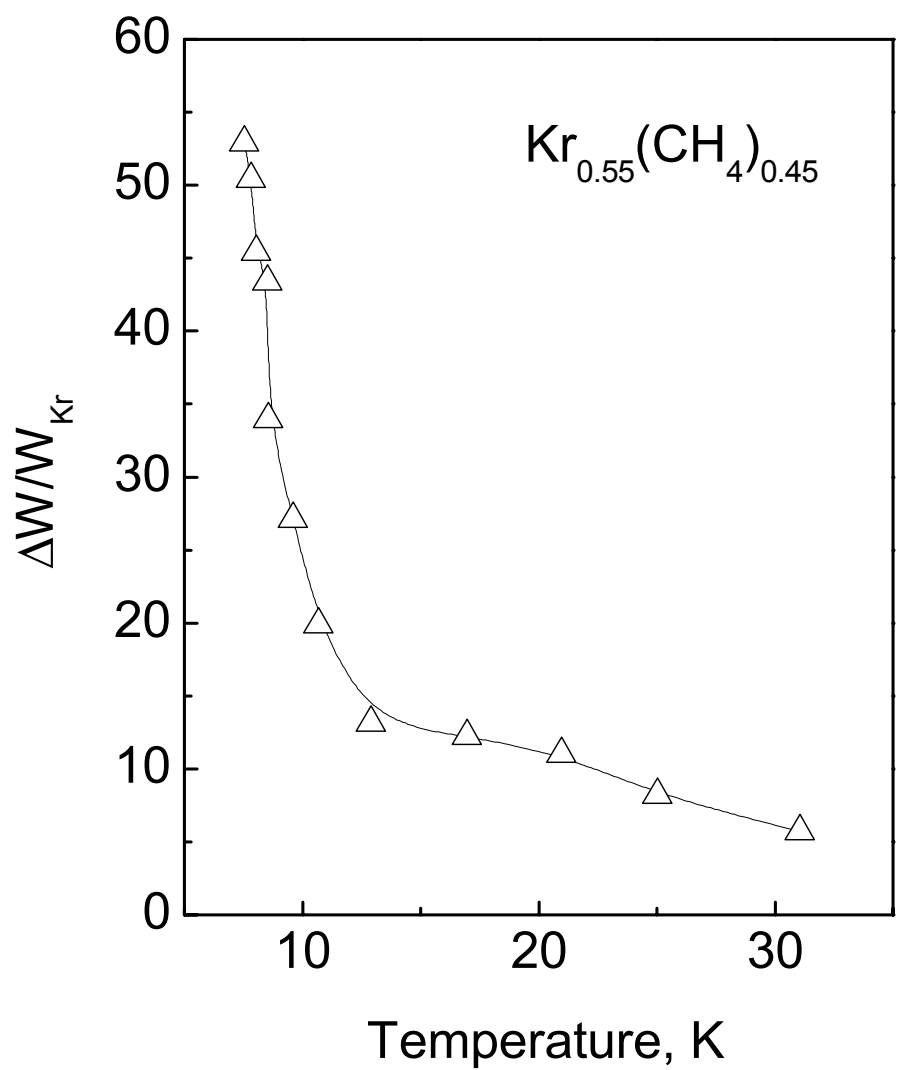

Fig. 3

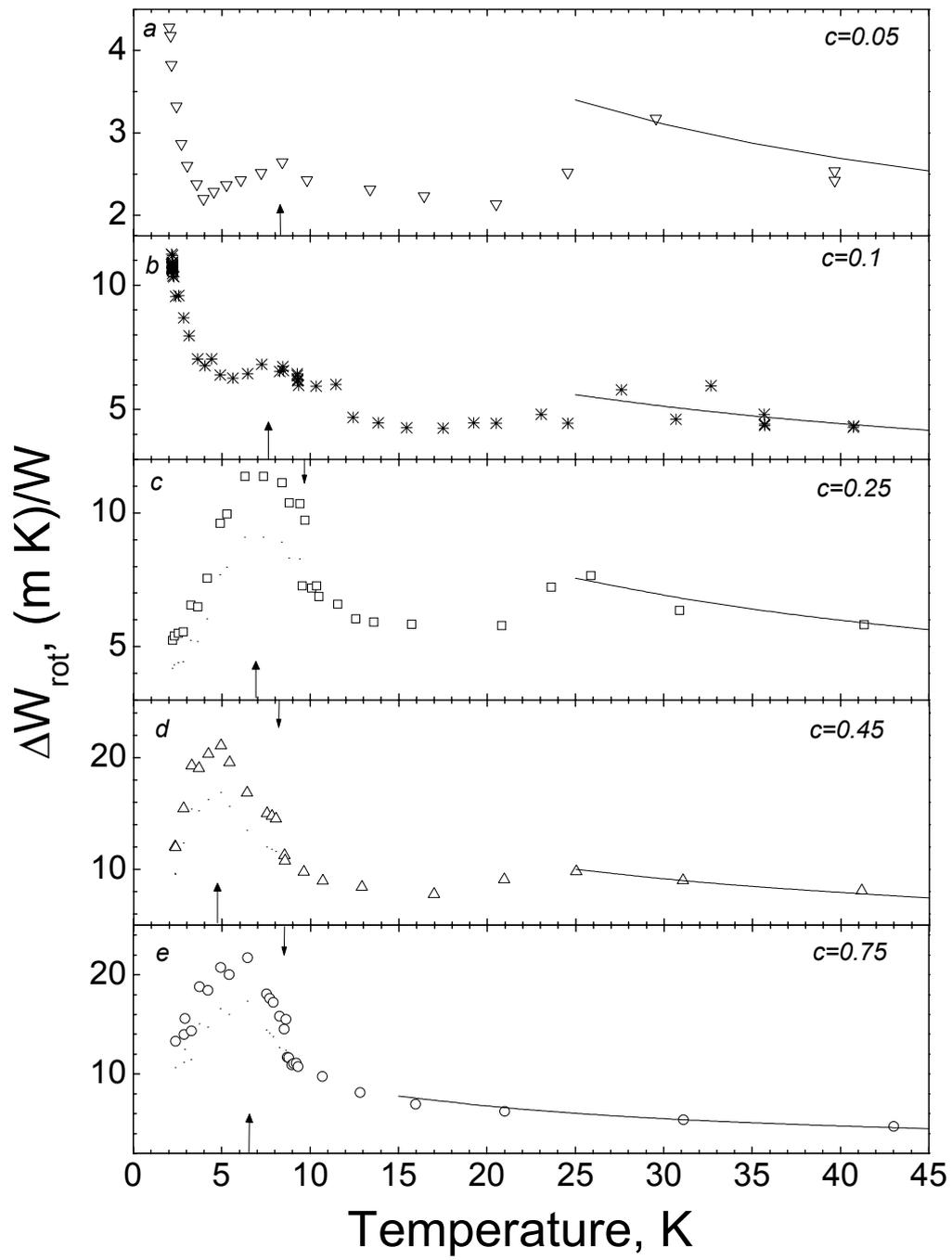

Fig. 4